\documentstyle[12pt,epsf]{article}  
\newcommand{\be}{\begin{equation}}
\newcommand{\ee}{\end{equation}}
\newcommand{\ben}{\begin{eqnarray}\displaystyle}
\newcommand{\een}{\end{eqnarray}}
\newcommand{\refb}[1]{(\ref{#1})}

\begin{document}

{}~ \hfill\vbox{\hbox{hep-th/0009246}
\hbox{CTP-MIT-3027}
\hbox{UUITP-09/00}}\break 

\vskip 3.0cm

\centerline{\large \bf  Effective Tachyon Dynamics in Superstring Theory}
\vspace*{1.5ex}

\vspace*{8.0ex}

\centerline{\large \rm Joseph A. Minahan\footnote{E-mail: minahan@mit.edu}}
\vspace*{2.5ex}
\centerline{\large \it Department of Theoretical Physics}
\centerline{\large \it Box 803, SE-751 08 Uppsala, Sweden}
\vspace*{3.0ex}
 \centerline{\large \rm and}
\vspace*{3.0ex}
 \centerline{\large \rm Barton Zwiebach\footnote{
E-mail: zwiebach@mitlns.mit.edu}}

\vspace*{2.5ex}

\centerline{\large \it Center for Theoretical Physics}
\centerline{\large \it Massachusetts Institute of Technology}
\centerline{\large \it  Cambridge, MA 02139, USA}

\vspace*{4.5ex}
\medskip
\centerline {\bf Abstract}

\bigskip
A recently proposed $\ell=\infty$ field theory model of tachyon 
dynamics for unstable bosonic D-branes has been shown to arise 
as the two-derivative truncation of (boundary)-string field theory. 
Using an $\ell\to \infty$ limit appropriate to stable kinks we obtain 
a  model for the tachyon dynamics on  unstable D-branes or D-brane 
anti-D-brane pairs of superstring theory.  The tachyon potential 
is a positive definite even function of the tachyon, and at the 
stable global minima there is no on-shell dynamics.  The kink solution 
mimics nicely the properties of stable D-branes: the spectrum of the 
kink consists of infinite levels starting  at zero mass, with spacing 
double the value of the tachyon mass-squared.  It is natural to expect 
that this model will arise in (boundary) superstring field theory.
\vfill \eject
\baselineskip=17pt

Field theory models of tachyon dynamics have been
a useful tool to understand the realization of 
Sen's conjectures on tachyon condensation and D-brane annihilation 
\cite{senconj}.  
The p-adic string  models \cite{0003278}, defined
with a choice of prime number, describe tachyon dynamics with
infinitely many spacetime derivatives and correctly show the dissappearance
of on-shell dynamics at the stable vacuum. Lumps of any codimension
can be found exactly.  An $\ell=3$ model 
\cite{0008227} is a solvable model with lump solutions whose
worldvolume theory can be calculated exactly, and where tachyon
condensation proceeds in a way mathematically similar to the
case in open string field theory. The lump spectrum contains a
continuous sector, somewhat reminiscent of the closed string sector.
An $\ell=\infty$ model \cite{0008231}
combines solvability with strikingly 
stringy properties. The spectrum of
the lump solutions representing unstable D-branes contains a tachyon 
of the correct mass, and equally spaced infinite levels. There is no continuous
spectrum, and nothing survives after tachyon condensation. This $\ell=\infty$
model can also be obtained as the $p\to 1$ limit in p-adic string theory 
\cite{0009103}. In addition, the $\ell=\infty$ model can be supplemented
with gauge field dynamics also showing stringy properties \cite{0008231}.

It was recently pointed out \cite{0009103,0009148} that
the $\ell=\infty$ model arises as a two-derivative reduction
of (boundary) string field theory (B-SFT). In particular, 
the tachyon potential in the model is the exact potential.
B-SFT is a formally background independent approach to
string field theory proposed in \cite{9208027}
and developed in \cite{9210065,9303067,9303143}.
As opposed to cubic string field theory \cite{WITTENBSFT}, B-SFT 
is not precisely defined in general, but can be defined concretely for particular
families of backgrounds, notably along
a set of backgrounds related by relevant tachyonic 
deformations \cite{9303067}.\footnote{For related computations in
boundary CFT
see \cite{list}.} Indeed, the works of \cite{0009103,0009148},
supplemented with
the normalization  calculation of \cite{0009191}, provide an exact verification
of the energetics  aspect of the tachyon conjectures. While strong
evidence for
the  conjectures has been obtained using cubic string field theory
\cite{cubiccalc}  (and superstring field theory \cite{superchecks}
for the analogous superstring conjectures) exact
verification in this approach appears to require further understanding
of 
the properties of the star algebra of open strings \cite{insight}.

In this paper, following \cite{0008231}, we will produce a 
(two-derivative) field theory model
for tachyon dynamics in superstring theory. Again, we will
be guided by the requirements of solvability of the model, absence
of dynamics on the spatially homogeneous stable vacuum, and 
the demand that the (stable) kink arising from the model, representing
a stable D-brane, should have a string like spectrum on its worldvolume.
The resulting model is
\be
\label{themodel}
S = - 8{\cal T} \int dt\,d^px \,\Bigl( \,\,{1\over 2}\, 
e^{-2T^2} \,\partial_\mu T \,\partial^\mu T  + {1\over 8} 
\, e^{-2T^2} \,\Bigr)\,.
\ee
Here we are considering the tachyon field theory on the
world volume of a non-BPS Dp-brane of superstring theory.
It is then natural to extend this action to 
a coincident 
Dp-brane and anti-Dp-brane pair, 
each separately BPS. 
Since the tachyon field on the
Dp-brane anti-Dp-brane pair is complex, in this case one must
replace $T^2 \to TT^*$  and $(\partial T)^2\to \partial T^* \,\partial
T$.  The tension ${\cal T}$ must be adjusted accordingly. 
The following points should be noted:
\begin{itemize}

\item The unstable vacuum is $T=0$, where the tachyon
mass squared is $M_T^2 = -1/2$, in units where $\alpha'=1$.

\item The potential is positive definite and an even  function
of $T$. 

\item The stable minima are
at $T= \pm \infty$.  At these points the effective mass squared
of the tachyon is infinite. 

\item The non-BPS tachyon dynamics of \refb{themodel} admits a
kink (and anti-kink)
solution  describing a codimension one stable D-brane localized 
along an $x$ coordinate.
The profile is given simply by  $T = \pm x/2$.

\item The spectrum on the kink will be shown to consist of equally
spaced mass
levels $M^2 =0,1,2, \cdots$. Notice that the spacing is twice the
value of $|M_T^2|$, as in string theory. 

\item The tension ${\cal T}_{kink}$ satisfies the relation
$ {\sqrt{2}\over 2\pi} \, {{\cal T}_{kink}\over {\cal T}} 
= {2\over \sqrt{\pi}} \simeq 1.128$.  
In string theory this ratio takes the value of unity.

\end{itemize}
The fact that \refb{themodel} satisfies the above properties
leads us to conjecture that it will arise as the two derivative
reduction of the yet to be analyzed (boundary) superstring 
field theory. In particular, the potential should be the 
exact one.  

\bigskip
\noindent
\underbar{Constructing the model}.~
Recall how the various $\ell$ models were constructed
\cite{GJa,0008231}. The
idea was finding a field theory admiting
a stationary solution whose spectrum is governed by
the Schroedinger problem with reflectionless potential\footnote{For a
pedagogical   review on these and other solvable Hamiltonians with references
to the early literature see \cite{9405029}.  Applications of 
reflectionless systems to fermions can be found in \cite{9408120}.} 
$U_\ell (x)   = - \ell ( \ell + 1) \hbox{sech}^2 x$. For 
unstable lumps the ground state of the Schroedinger problem
gives rise to the tachyon on the lump. The next level
is associated with the translation mode of the lump, namely
it equals the derivative of the profile. Once the profile
is known one can readily reconstruct the potential that gives 
rise to it.  This time, since we want a stable lump, we
must identify the derivative of the profile with the ground
state wavefunction. Indeed, this familiar procedure
leads to the sine-Gordon soliton
and the $\phi^4$ kink solution for
the cases $\ell=1$ and $\ell=2$ respectively\cite{christ}. For higher
values of $\ell$ direct application of this
procedure leads to complicated
field theory potentials \cite{christ,GJ}. We will focus here
on the special case of interest $\ell\to\infty$, and will
see that with the help of a field redefinition the action
can be given in simple form. The general aspects of this field
redefinition, which could be used for arbitrary $\ell$
will be noted afterwords. 
\medskip

\begin{figure}[!ht]
\leavevmode
\begin{center}
\epsfbox{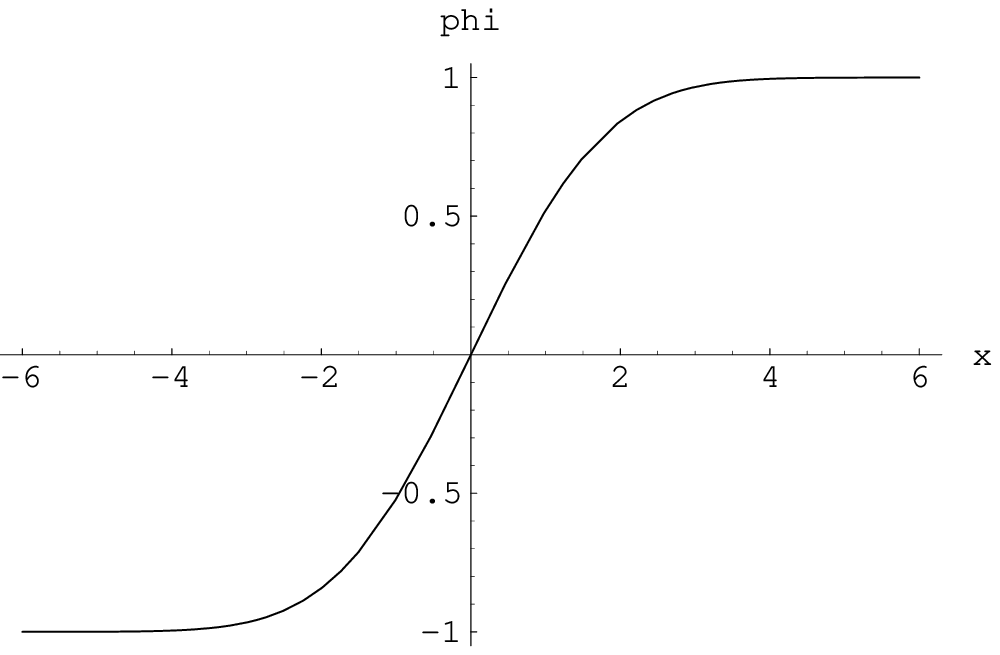}
\end{center}
\caption[]{\small The profile $\overline\phi (x) = \hbox{erf} (x/2)$
of the stable kink representing a D-brane. In the field variable
$T$ used in \refb{themodel} we have $\overline T (x) = x/2$.} \label{f1}
\end{figure}

We are looking for a model taking the form
\be
\label{fe0}
S \sim  - \int dt d^px \, \Bigl( {1\over 2} \,\partial_\mu \phi \,\partial^\mu
\phi + V(\phi) \Bigr)\,.
\ee
In the $\ell\to\infty$ limit of \cite{0008231} the ground 
state wavefunction becomes a gaussian $\sim \exp (-x^2/4)$.
We therefore identify this wavefunction with the spatial
derivative of the profile $\overline\phi (x)$ of the stable
lump solution of the model to be found.  Thus, we set
\be
\label{fe1}
{\overline\phi}' (x) = {1\over \sqrt{\pi}}\, \exp \Bigl( -{x^2\over
4}\Bigr) \,,
\ee
where the constants have been chosen for convenience.
It follows by integration that 
\be
\label{fe2}
\overline \phi (x) = \hbox{erf} \,\Bigl( {x\over 2}\Bigr)\,,\quad 
\hbox{erf} (x) \equiv {2\over \sqrt{\pi}} \int_0^x e^{-u^2} du \,,
\ee
where `erf' is the familiar error function. A plot of this profile is
shown in Fig.~\ref{f1}. It is a kink.

{}From the
equation of motion for the soliton 
\be
\label{fe3}
{1\over 2} ({\overline\phi}' (x))^2 = V (\overline \phi (x)) \,,
\ee
and from equations \refb{fe1}  and \refb{fe2} it follows  that 
\be
\label{fe4}
V (\overline \phi (x)) = {1\over 2\pi} \exp \Bigl( -{x^2\over 2}\Bigr)
= {1\over 2\pi} \exp \Bigl( -2 [\,\hbox{erf}^{-1} (\overline \phi)]^2 \Bigr)\,.
\ee
In the above erf${}^{-1}$ denotes the inverse function to erf. While
the above expression is implicit, it can be used,  
for example, to find an expansion for small $\phi$:
\be
\label{expsm}
V(\phi) = {1\over 2\pi} - {1\over 4} \phi^2 + {\pi\over 48} \phi^4 + \cdots
\,.
\ee
The mass squared of the tachyon field in the field theory model
is readily recognized to be $M_T^2 = -1/2$.
The fluctuation spectrum is also easily obtained. Recall that the 
Schroedinger potential for the fluctuations 
is $V''(\overline \phi (x))$ 
 and that $V'(\overline\phi(x))$ satisfies
\be
\label{fe3c}
\overline\phi''(x)=V'(\overline\phi(x)).
\ee
Taking a derivative on each side of \refb{fe3c}, we have
\be
\label{fe3d}
\overline\phi'''(x)=V''(\overline\phi(x))\overline\phi'(x).
\ee
But by construction, $\overline\phi'(x)$ is the ground state solution to
the harmonic oscillator equation.  Thus we have
\be
\label{schpot}
V''(\overline \phi (x)) = 
 -{1\over 2}\, + {x^2\over 4}\,.
\ee
Note that 
this also gives us an easy way of computing $V''(0)$, and hence the mass of the
tachyon in the open string vacuum, without actually having to invert the
error function.

Thus we have obtained (once more) the 
potential for the simple harmonic oscillator. The Schroedinger
equation determining the masses $m^2$ of the modes living on
the kink is 
\bigskip
\be
\label{fe5}
-\frac{d^2}{dx^2}\psi(x)+\left(-{1\over 2}+
{x^2\over 4}\right)\psi(x)=m^2\psi(x)\, ,
\ee
and the masses of the fields are therefore
\be
\label{masss}
m^2=n\,,  \quad n\ge0 \,.
\ee
Thus, we get a massless field and equally spaced massive fields. 
The spacing between adjacent mass levels 
is twice the value of $|M_T^2|$.
Note also that when $\overline\phi \to \pm 1$, $x \to \pm\infty$,
and $V''(\overline\phi)\to +\infty$. This
is the statement that the tachyon acquires infinite mass
on the vacuum $\phi = \pm 1$.

\medskip
The nature of the model is better appreciated after a 
field redefinition. Let 
\be
\label{redef}
\phi = \hbox{erf}\, (T) \quad \to \quad \partial \phi = {2\over \sqrt{\pi}}
\, e^{-T^2} \, \partial T \,.
\ee
Since erf${}^{-1} (\phi) = T$ we also have from \refb{fe4} that
\be
V(T) = {1\over 2\pi} e^{-2T^2}\,.
\ee
The kinetic term in the $T$ variable reads
\be
{4\over \pi} e^{-2T^2} (\partial T)^2 \,.
\ee
We have therefore recovered the action given in \refb{themodel}.
Finally, since $T= $erf${}^{-1} (\phi)$, the profile $\overline T(x)$
for the kink follows directly from  \refb{fe2} and is given as
\be
\overline T(x) = {x\over 2}\,.
\ee
Evaluation of the kink energy using the action \refb{themodel} gives
\be
\label{evenerg}
E =  8{\cal T} \int \,d^{p-1}y dx \,\Bigl( \,\,{1\over 2}\, 
e^{-2\overline T^2} \,\partial_x \overline T \,\partial_x \overline T  + {1\over
8} 
\, e^{-2\overline T^2} \,\Bigr)\, = 2 \sqrt{2\pi}\, {\cal T}\, \Bigl(\int
\,d^{p-1}y\Bigr)\,,
\ee
giving ${\cal T}_{kink} = 2 \sqrt{2\pi}\, {\cal T}$.  We therefore have 
\be
\label{quotient}
{\sqrt{2}\over 2\pi} \, {{\cal T}_{kink}\over {\cal T}} 
= {2\over \sqrt{\pi}} \simeq 1.128 \,.
\ee  
In string theory this ratio takes the value of unity\footnote{
Recall that the tension for a non-BPS Type IIA 
D$p$ brane has an extra factor of 
$\sqrt{2}$ as compared to the tension for a Type IIB BPS D$p$ brane.}.

The results found for  (boundary) bosonic SFT \cite{0009103,0009148},
together with the
above results suggest that in (boundary) superstring field theory
one may use tachyonic backgrounds of the type $T(X) = a + u X$. 
The vacuum with no brane would be at
$(a, u) = (\infty, 0)$ and the kink solution would correspond 
to $(a, u) = (0, \infty)$. For the effective model, the kink is obtained
with finite $u$, namely $u=1/2$. It is therefore pleasing to see that
the tension of the kink, as estimated in the model, is larger than
the correct value in string theory, though, happily, not by much.
It is also satisfying to see that the ratio in \refb{quotient} would have
been less than 1, were it not for the extra factor of $\sqrt{2}$ arising
from the non-BPS brane tension. 

\bigskip
\noindent
\underbar{Reconstructing Potentials from Profiles}.  As one can
see from the above discussion, the reconstruction of the field
theory from the kink profile is aided by a field redefinition. 
This field  redefinition makes the kinetic term more complicated but allows
a closed form solution.  In general, assume one is given 
a kink profile 
\be
\overline \phi (x) = {\cal P}(x) ,
\ee
where  ${\cal P}(x)$ is  
either a monotonically increasing or 
a monotonically decreasing function of the coordinate $x \in [-\infty, \infty]$,
as appropriate for a kink solution. It then
follows that 
\be
\label{fe3a}
 V (\overline \phi (x)) ={1\over 2} ({\cal P}' (x))^2  \,.
\ee
Introduce now a new field $T$ via
\be 
\phi =   {\cal P}(T) \, \quad\to \quad   T = {\cal P}^{-1}(\phi)\,.
\ee
With this field variable the profile of the kink is just
$\overline T (x) = x$. 
The monotonicity of ${\cal P}$ ensures that the relation between
$\phi$ and $T$ is single valued, with $T$ defined on the full
real line. Then 
\be
\label{fe3b}
 V (\overline \phi (x)) ={1\over 2} \Bigl({\cal P}' (\,{\cal P}^{-1}[\,\overline
\phi\,]\,)\Bigr)^2 ={1\over 2} \Bigl({\cal P}' (\overline T)\Bigr)^2 \,,
\ee
showing that the potential is simple in terms of the new field variable.
Since $\partial \phi = {\cal P}'(T) \partial T$ the final model is just
\be
\label{finmod}
S = - \int dt dx \Bigl({\cal P}' ( T)\Bigr)^2 \,
\Bigl(\, {1\over 2} (\partial T)^2  + {1\over 2} \,\Bigr)\,.
\ee

\bigskip
\noindent {\bf Acknowledgments}:
B. Z. is happy to thank H. Verlinde for a stimulating discussion
that rekindled his interest in this problem. We thank
J. Goldstone,  R. Jackiw, and A. Sen for discussions. We
are particularly indebted to R.~Jaffe for his useful input
and insight. 
This work was
supported in part by DOE contract \#DE-FC02-94ER40818.

\end{document}